\documentclass[12pt,twoside]{article}

\evensidemargin=\oddsidemargin
\def\Title#1#2#3{%
    \baselineskip=18pt
    \begin{center}
          {\large\bf{#1} \\ }
          \bigskip\bigskip
          {#2} \\
          {#3} \\
    \end{center}}
\long\def\Abstract#1{%
         \bigskip
         \parbox{0.93\textwidth}{%
                 \begin{center}
                       {\bf Abstract} \\
                 \end{center}
                 \medskip{\baselineskip=14pt #1}
                 \vss}
         \bigskip}

\begin{document}

\Title{The formulation of General Relativity\\
in extended phase space\\
as a way to its quantization}%
{T. P. Shestakova}%
{Department of Theoretical and Computational Physics,
Southern Federal University\footnote{former Rostov State University},\\
Sorge St. 5, Rostov-on-Don 344090, Russia \\
E-mail: {\tt shestakova@sfedu.ru}}

\Abstract{Our attempts to find an explanation for quantum behavior of the Early Universe appeal, as a rule, to the Wheeler -- DeWitt Quantum Geometrodynamics which relies upon Hamiltonian formulation of General Relativity proposed by Arnowitt, Deser and Misner (ADM). In spite of the fact that the basic ideas of this approach were put forward about fifty years ago, even now we do not have clear understanding what Hamiltonian formulation of General Relativity must be. An evidence for it gives a recent paper by Kiriushcheva and Kuzmin [gr-qc/0809.0097], where the authors claim that the formulation by ADM and that by Dirac made in his seminal work of 1958 are not equivalent. If so, we face the question what formalism should be chosen. Another problem is that we need a well-grounded procedure of constructing a generator of transformations in phase space for all gravitational variables including gauge ones. It suggests the notion of extended phase space. After analyzing the situation, we show that Hamiltonian formulation in extended phase space is a real alternative to Dirac and ADM formulations and can be constructed to be equivalent to the original (Lagrangian) formulation of General Relativity. Quantization in extended phase space is straightforward and leads to a new description of quantum Universe in which an essential place is given to gauge degrees of freedom.}

\bigskip

In 2008 fifty years passed after the publication of the Dirac famous paper, devoted to Hamiltonian form of the theory of gravitation \cite{Dirac}. However, the formulation proposed by Arnowitt, Deser and Misner (ADM) may have become even more recognized and gave a basis of the Wheeler -- DeWitt Quantum Geometrodynamics \cite{ADM}.

Meanwhile, in the same year, 2008, there appeared the paper by Kiriushcheva and Kuzmin \cite{KK}, where the authors claim that the formulation by Dirac and that by ADM are not equivalent since these two formulations are related by a non-canonical transformation of phase space variables. I would like to emphasize that the authors of this paper have raised a problem: Should we abandon the ADM parametrization (and also any others) if the new variables are not related with the old ones by a canonical transformation? We come to the conclusion that {\it even now, fifty years after the Dirac paper, we are not sure what formulation of Hamiltonian dynamics for General Relativity is correct}. Do we have any correct formulation?

We do not have strict mathematical rules how to construct Hamiltonian dynamics for a theory with constraints. The rules, proposed originally by Dirac, were then modified by many authors. In particular, we need a well-grounded
procedure of constructing a generator of transformations in phase space for all gravitational variables including gauge ones. According to Dirac, the constraints or their linear combinations play the role of generators of gauge transformations. However, the constraints as generators cannot produce correct transformation for the $g_{0\mu}$ components of metric tensor or the lapse and shift functions. To avoid this difficulty, some other algorithms were suggested to construct the generator of transformations, see, for example, \cite{Cast,BRR}. The most of the methods proposed rely upon the algebra of constraints that is not invariant under the choice of parametrization.

One can say that the difference in the algebra of constraints is a consequence of the fact that the two parametrizations are not related by a canonical transformation. However, {\it such a transformation has to involve gauge variables which in the original Dirac approach played the role of Lagrangian multipliers at constraints and are not included into the set of canonical variables}. To treat these variables on the equal basis with the others, {\it one should extend the phase space}.

The idea of extended phase space appeared in the works by Batalin, Fradkin and Vilkovisky (BFV) \cite{BFV1,BFV2,BFV3} where their approach to path integral quantization of gauge theories was proposed. In their approach the generator (BRST charge) also depends on the algebra of constraints. This makes us search for another way of constructing Hamiltonian dynamics in extended phase space. Such a way has been proposed in our papers \cite{SSV1,SSV2}.

Consider an isotropic model with the Lagrangian
\begin{equation}
\label{Lagr1}
L=-\frac12\frac{a\dot a^2}N+\frac12 Na
\end{equation}
We introduce the missing velocities into the Lagrangian by means of gauge conditions in differential form. The condition $N=f(a)$ gives $\dot N=\displaystyle\frac{df}{da}\;\dot a$. One should also include the ghost sector into the model, that leads to the full Lagrangian
\begin{equation}
\label{Lagr3}
L=-\frac12\frac{a\dot a^2}N+\frac12 Na
 +\lambda\left(\dot N-\frac{df}{da}\;\dot a\right)
 +\dot{\bar\theta}\left(\dot N-\frac{df}{da}\;\dot a\right)\theta
 +\dot{\bar\theta}N\dot\theta.
\end{equation}
The conjugate momenta are:
\begin{equation}
\label{mom1}
\pi=\lambda+\dot{\bar\theta}\theta;\quad
p=-\frac{a\dot a}N-\pi\frac{df}{da};\quad
\bar{\cal P}=N\dot{\bar\theta};\quad
{\cal P}=N\dot\theta.
\end{equation}

Then we go to a new variable, $N=v(\tilde N,\; a)$ while the rest variables being unchanged, $a=\tilde a$, $\theta=\tilde\theta$, $\bar\theta=\tilde{\bar\theta}$. It is the analog of the transformation from the original gravitational variables $g_{\mu\nu}$ to the ADM variables. Indeed, in the both cases only gauge variables are transformed. It was shown in \cite{KK} that such a transformation is not canonical. The reason is that the momenta conjugate to physical variables also remained unchanged. The situation in extended phase space is different. After going to the new gauge variable the Lagrangian is written as
\begin{equation}
\label{Lagr4}
L=-\frac12\frac{a\dot a^2}{v(\tilde N,\; a)}
 +\frac12\;v(\tilde N,\; a)a
 +\pi\left(\frac{\partial v}{\partial\tilde N}\;\dot{\tilde N}
  +\frac{\partial v}{\partial a}\;\dot a
  -\frac{df}{da}\;\dot a\right)
 +v(\tilde N,\; a)\dot{\bar\theta}\dot\theta.
\end{equation}
The new momenta are related with the old ones as following:
\begin{equation}
\label{mom2}
\tilde\pi=\pi\frac{\partial v}{\partial\tilde N};\quad
\tilde p=p+\pi\frac{\partial v}{\partial a};\quad
\tilde{\bar{\cal P}}=\bar{\cal P};\quad
\tilde{\cal P}={\cal P}.
\end{equation}

It is easy to demonstrate that now the transformation in extended phase space is canonical. In particular, the Poisson brackets among all phase variables maintain their canonical form. Moreover, the existence of global BRST invariance enables us to construct the generator of transformations in extended phase space making use of the first Noether theorem. The BRST generator can be constructed if the theory is not degenerate, i.e. derivatives of the Lagrangian with respect to all velocities are not zero, and the Lagrangian and Hamiltonian dynamics are completely equivalent. The first condition is guaranteed by the extension of phase space, the second one is ensured by construction of the Hamiltonian dynamics itself presented in our papers \cite{SSV1,SSV2}.

In the extended phase space approach we do not need to abandon generally accepted rules of constructing a Hamiltonian form of the theory or invent some new rules. Indeed, in our approach
\begin{itemize}
\item the Hamiltonian is built up according to the usual rule $H=p_a\dot q^a-L$;
\item the Hamiltonian equations in extended phase space are completely equivalent to the Lagrangian equations;
\item due to global BRST invariance it appears to be possible to construct the BRST charge in conformity with
the first Noether theorem which produces correct transformations for all phase variables.
\end{itemize}

Dirac was not tired of repeating that ``any dynamical theory must first be put in the Hamiltonian form before one can quantize it'' \cite{Dirac}. The Hamiltonian formulation in extended phase space has been proved to be a real alternative to Dirac and ADM formulations. Quantization in extended phase space is straightforward and leads to a new description of quantum Universe in which an essential place is given to gauge degrees of freedom: a gravitating system is described from a viewpoint of reference frame from which it can be observed.

\section*{Acknowledgments}
I would like to thank Giovanni Montani and Francesco Cianfrani for attracting my attention to the paper \cite{KK}. I am grateful to the MG12 Organizers for financial support that let me take part in the MG12 Meeting. My participation in the Meeting was also partially supported by the RFBR grant 09-02-08224.

\end{document}